\documentclass[%
 reprint,
 superscriptaddress,
 amsmath,amssymb,
 aps,
floatfix,
]{revtex4-1}

\usepackage{graphicx}
\usepackage{dcolumn}
\usepackage{bm}
\usepackage{braket}
\usepackage{color}

\newcommand{\affA}{%
\affiliation{
     National Institute of Information and Communications Technology
     (NICT), \\
     4-2-1 Nukui-kitamachi, Koganei, Tokyo 184-8795, Japan}
     }
\newcommand{\affB}{%
\affiliation{
	Department of Physics, 
	Gakushuin University, 
	1-5-1 Mejiro, Toshima-ku, Tokyo 171-8588, Japan}
	}
\newcommand{\affC}{%
\affiliation{
    Department of Engineering and Applied Sciences, 
	Sophia University, \\
	7-1 Kioi-cho, Chiyoda-ku, Tokyo 102-8554, Japan}
	}
\newcommand{\affD}{%
\affiliation{
    National Institute of Advanced Industrial Science and Technology 
	(AIST), \\
	1-1-1 Umezono, Tsukuba, Ibaraki 305-8568, Japan}
	}
\newcommand{\affE}{%
\affiliation{
    Quantum Information Processing Group, 
	Raytheon BBN Technologies, \\
	10 Moulton Street, Cambridge, MA 02140, USA}
	}

\begin{document}


\title{Ultrabroadband direct detection of 
nonclassical photon statistics at telecom wavelength}
\date{\today}

\author{Kentaro Wakui}
\affA%
\author{Yujiro Eto}
\affA\affB%
\author{Hugo Benichi}
\affA%
\author{Tetsufumi Yanagida}
\affA\affC%
\author{Shuro Izumi}
\affA\affC%
\author{Kazuhiro Ema}
\affC%
\author{Takayuki Numata}
\affD%
\author{Daiji Fukuda}
\affD%
\author{Masahiro Takeoka}%
\affA\affE%
\author{Masahide Sasaki}%
\affA%

\begin{abstract}

Broadband light sources play essential roles in diverse fields, 
such as high-capacity optical communications,
optical coherence tomography, optical spectroscopy, 
and spectrograph calibration.
Though an ultrabroadband nonclassical state from standard spontaneous parametric down-conversion may serve as a quantum counterpart, 
its detection and quantum characterization have been a challenging task. 
Here we demonstrate the quantitative characterization of a multimode structure 
in such an ultrabroadband (150\,nm FWHM) squeezed state at telecom wavelength (1.5\,$\mu$m). 
The nonclassical photon distribution of our highly multimode state is directly observed using a superconducting transition-edge sensor. 
From the observed photon correlation functions, 
we show that several tens of different squeezers are coexisting in the same spatial mode. 
We anticipate our results and technique open up a new possibility to generate and characterize 
nonclassical light sources for a large-scale optical quantum network in the frequency domain.

\end{abstract}


\maketitle

Generating and characterizing nonclassical light fields, 
are indispensable prerequisites 
for optical quantum information processing (QIP)
\cite{KLM01, O'Brien09, Weedbrook12}. 
Spontaneous parametric down-conversion (SPDC) 
in a nonlinear crystal 
is an established source of nonclassical light field,
such as entangled photons or squeezed states
\cite{MWbook95, GKbook05}.
Moreover, SPDC can easily generate extremely broadband nonclassical states
\cite{Okano12}.
Although frequencies in such an ultrabroadband state are strongly entangled
\cite{KR97, Grice01},
the state can be decomposed into separable broadband modes
under certain assumptions
\cite{ZC90, Law00, Sasaki_Suzuki_PRA73_2006, Rohde07}.
The decomposed multimode state 
may consist of more than hundreds of broadband modes
that is potentially useful for future high-capacity quantum communication 
network or large-scale quantum information processing.

Toward utilizing the full potential of the ultrabroadband multimode 
nonclassical light, the first challenge to be overcome is 
to characterize the nonclassical properties of such a huge quantum system. 
Though in principle one can fully reconstruct an $N$-mode optical state 
by the homodyne tomography 
\cite{Breitenbach97, Lvovsky09}, 
it appears not only technically challenging but extremely time-consuming 
simply because one has to prepare $N$ different measurements with 
different appropriate local oscillators (LOs) for highly multimode states,
especially when $N$ is large e.g. $N>100$.
So far, multimode analyses under reasonable assumptions have theoretically revealed that 
the ultrabroadband nonclassical light from SPDC may consist of a multimode squeezed state
\cite{U'Ren_JSD_thermal_dist_03, Christ11}. 
Nevertheless, the quantitative evaluation of such a huge multimode squeezed state has not been 
experimentally investigated to date even with the use of any assumptions.

In this paper, we demonstrate the quantum characterization 
of the ultrabroadband optical state from SPDC at telecom wavelength 
(centered at $1.5\,\mu$m, with an estimated bandwidth of 150\,nm) by an alternative very simple approach, 
an ultrabroadband photon-number-resolving detection (PNRD). 
We characterize the nonclassical property by the following two observations:
First, we directly observe the nonclassical photon statistics of a whole 
multimode state violating the Klyshko's criterion
\cite{Klyshko_original_96, Waks06}.
Second, by using the method in 
\cite{Christ11}, 
we demonstrate that our source contains several tens of 
independent squeezed vacua in the broadband mode basis 
and determine the degree of squeezing in each mode.
These are observed by simply detecting photon-numbers of the whole pulse.

Our study is motivated by several related works. 
The direct observation of the nonclassical photon statistics 
of the SPDC source was demonstrated with a visible light photon counter (VLPC) 
\cite{Waks04, Waks06} 
but only at the visible wavelength. 
The relation between the nonclassical photon statistics 
and the frequency or spatial multimode structure
in the SPDC sources was successfully observed in
\cite{Avenhaus08, Mauerer09}, 
which could be useful to characterize 
the multimode squeezed states from SPDC 
\cite{Christ11, Goldschmidt13}. 
In 
\cite{Avenhaus08, Mauerer09, Goldschmidt13}, 
multiple avalanche photodiodes (APDs) were used as a PNRD. 
However, the APD-based PNRD is not suitable for our purpose since its 
detection efficiency (DE) and the number of countable photons is 
severely limited.

Here we use a superconducting transition-edge sensor (TES) as an ultrabroadband PNRD. 
TES has the number resolution at higher photon numbers with DE close 
to 100\,\% and the sensitivity in an extremely wide spectral range 
including all the important optical wavelengths
\cite{Lita08,Fukuda09,Fukuda11,Gerrits11_OnChipTES}.
It thus overcomes all technical obstacles to detect ultrabroadband quantum states.
Specifically, with an adequate optical coating \cite{Fukuda09}, our TES covers the S-, C-, and L-band at telecom wavelength.

Unlike the tomographic approach, 
our method does not allow to reconstruct 
full information of quantum state. 
However, it drastically reduces the number of measurement 
(just measuring photon numbers) and still captures 
nontrivial quantum property of the state. 
We anticipate our results and techniques open up a new way to
fully use nonclassical resources in the frequency domain 
\cite{Menicucci08, Pysher11, Pinel12}. \\

Our light source is a single-beam squeezer  
in a collinear configuration  
where 
the pump, signal, and idler beams are in the same spatiotemporal 
and polarization mode. 
The generation process is mathematically described by 
the following unitary transformation 
\cite{Christ11, Lvovsky_Wasilewski_Banaszek2007JMO_Decomposing_OPA_Mode} 
\begin{eqnarray}
\hat{\bm{S}}
&=&
\exp 
\Biggl[ -\frac{i}{\hbar} 
\Biggl( 
A \int d\omega_{\mathrm{s}} \int d\omega_{\mathrm{i}} f
  \left( \omega_{\mathrm{s}}, \omega_{\mathrm{i}} \right) 
  \hat{a}^{\dagger} (\omega_{\mathrm{s}}) 
  \hat{a}^{\dagger} (\omega_{\mathrm{i}}) 
\nonumber\\
&&+ \mathrm{h.c.} 
\Biggr) 
\Biggr].
\end{eqnarray}
Here $A$ denotes the overall efficiency of the squeezing, 
$ f\left( \omega_{\mathrm{s}}, \omega_{\mathrm{i}} \right) $ 
is the joint-spectral distribution (JSD) 
in terms of the signal and idler angular frequencies, 
$\omega_{\mathrm{s}}$ and $\omega_{\mathrm{i}}$,
and
$ \hat{a}^{\dagger}_{\mathrm{s}} (\omega_{\mathrm{s}}) $ 
($ \hat{a}^{\dagger}_{\mathrm{i}} (\omega_{\mathrm{i}}) $)
represents the creation operator of the signal (idler) field 
in the continuous spectrum of 
$\omega_{\mathrm{s}}$ ($\omega_{\mathrm{i}}$). 
These continuous modes are, however, 
not necessarily a convenient representation to deal with 
the multimode characteristics, 
because they cannot be decoupled from each other 
for most of practical JSDs.  
For a JSD which distributes in an effectively finite range, 
one can find a more convenient discrete set of decoupled modes 
\cite{ZC90,Sasaki_Suzuki_PRA73_2006}.

In fact, when the JSD is engineered to be symmetric, 
which is the case here, 
it can then be decomposed into the following form 
\begin{equation}
\frac{i}{\hbar} 
A^{*} f^{*}\left( \omega_{\mathrm{s}}, \omega_{\mathrm{i}} \right) 
= \sum_k r_k \phi_k \left( \omega_{\mathrm{s}} \right) 
             \phi_k \left( \omega_{\mathrm{i}} \right).
\end{equation}
in terms of a complete orthonormal set
$\{ \phi_k (\omega) \}$ 
\cite{Christ11}. 
Note, this mode corresponds to the shape of the frequency distribution for signal and idler.
Here $r_k$ is a modal amplitude 
corresponding to a squeezing parameter for mode $k$. 
Introducing the field operators for the broadband basis modes
\cite{Rohde07},
\begin{equation}
\hat A_k
=
\int d\omega
\phi_k (\omega) \hat{a} (\omega)
\end{equation}
the above transformation can be expressed as 
\begin{equation}
\hat{\bm{S}}
=
\exp\left[
\sum_k r_k (\hat{A}_k^{\dagger 2}-\hat{A}_k^{ 2})\right].
\end{equation}

The generated state is a multimode squeezed vacuum state as  
\begin{eqnarray}
\ket{\bm{r}} = \hat{\bm{S}} \ket{0}.
\end{eqnarray}
Thus this state only includes even number of photons. 
In practice, however, 
inevitable losses in the generation, propagation, and detection processes 
cause the vacuum invasion, 
making the pure state $\ket{\bm{r}}$ a mixed one, 
whose statistics has odd number components as well. 
If losses are below a certain level, 
one can still observe even-rich photon-number statistics. 
Direct observation of this even-odd number oscillation 
and 
multimode analysis on the distribution of $r_k$ 
are our tasks here.

\begin{widetext}
\begin{figure*}[ht]
\centerline{\includegraphics[width=15cm]{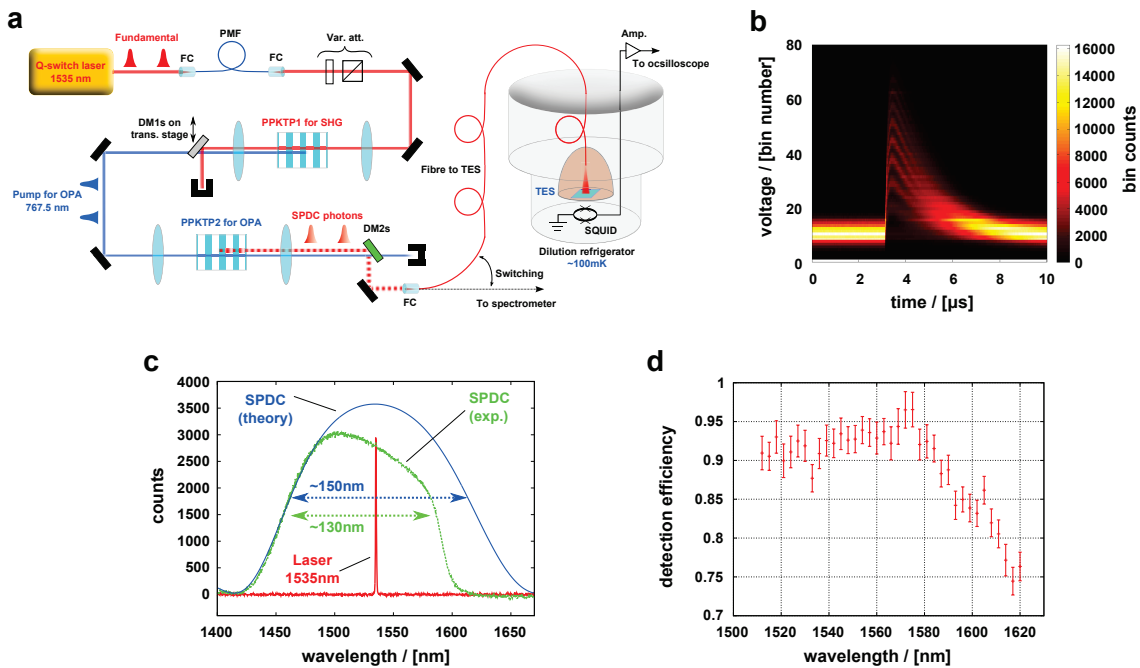}}
\caption{\textbf{Experimental setup.}
(\textbf{a}) 
Schematic diagram of the experimental setup. 
The fundamental light is guided into 
a polarization-maintaining fiber (PMF, used as a spatial-mode filter)
via a fiber-coupler (FC), 
passes through variable attenuator (Var. att.) 
consisting of a half wave plate and a polarizing beam splitter, 
and is then injected into a PPKTP crystal for SHG. 
The SHG output pumps another PPKTP crystal for squeezing. 
DM1s (DM2s) are dichroic mirrors 
acting as a high reflector for the fundamental (pump) light 
and as a high transmitter for the pump (fundamental) light.
SPDC output is coupled into an optical fiber 
and guided to a TES or a spectrometer. 
(\textbf{b}) 
Typical waveforms from the TES. 
Total $1\times10^{7}$ waveforms are overlaid. 
The vertical direction is divided into bins 
and voltage counts are summed with respect to each bin.
Each waveform contains 1000 points (horizontal direction). 
(\textbf{c}) 
Spectrum of the SPDC. 
The red line shows spectrum of the fundamental laser source. 
The green dotted and blue lines are 
measured and theoretical spectra of SPDC, respectively. 
(\textbf{d}) 
Wavelength dependence of TES's detection efficiency 
from 1512\,nm to 1620\,nm.
}
\label{setup1}
\end{figure*}
\end{widetext}

Figure \ref{setup1}a shows the experimental setup. 
The fundamental light source is a single-longitudinal-mode, passively Q-switched, 
diode-pumped solid-state laser (Cobolt, Tango), 
operating at 1535\,nm with a pulse width of 4\,ns 
and a repetition rate of 3.1\,kHz. 
This is used as the fundamental light, 
and is injected into 
a 10\,mm long, periodically poled KTiOPO${}_4$ (PPKTP) crystal 
for second harmonic generation (SHG) at 767.5\,nm. 
This second harmonic light is then used as 
a pump for squeezing. 
An unconverted fundamental light after SHG 
is used as a guiding beam for fiber-coupling to TES, 
but effectively eliminated by dichroic mirrors 
when photon counting experiment is implemented.
The pump is guided into the second type-0 PPKTP crystal 
employed as the squeezer.
The PPKTP crystals for SHG and the squeezer are mounted on copper stages 
and temperature-stabilized.
The multimode squeezed vacuum state is fiber-coupled 
and guided to the TES or to a spectrometer. 
Output waveform from the TES is amplified by a SQUID device 
and room-temperature electronics (Magnicon).
Each waveform was sampled 
using a digital oscilloscope (DPO7104, Tektronix), 
and sent to a host computer for noise filtering and data analysis.

Figure \ref{setup1}b shows raw waveforms from the TES.
Total $1\times10^{7}$ waveforms are stored and overlaid here. 
The red line in Fig. \ref{setup1}c 
is a spectrum of the fundamental light, 
whose linewidth is $\sim$0.05\,nm. 
The green dotted line is 
a spectrum of the SPDC
measured with the spectrometer (Acton) 
for a 30\,sec accumulation time.
Measured full-width at half maximum (FWHM) is roughly 130\,nm. 
The blue line is a theoretically derived spectrum 
using the present experimental parameters (see Appendix \ref{AP3_spectrum}), 
which is symmetric about the fundamental light spectrum. 
The asymmetry seen in the measured spectrum (green dotted line) 
is an artifact caused by the DE fall-off 
of an InGaAs photo detector in the spectrometer. 
Actually the spectrometer DE gradually drops when the wavelength exceeds 1500\,nm.
Then the DE starts steeply falling at around 1580\,nm 
and reaches almost 0\% at around 1650\,nm.
The blue theoretical spectrum captures  
the actual SPDC spectrum 
whose FWHM extends over 
almost 150\,nm in the telecom window, 
covering the S-, C-, and L-band. 

The wavelength dependence of our TES's DE is shown 
in Fig. \ref{setup1}d (also see Appendix \ref{AP1_DE_meas}). 
It is almost flat from 1512\,nm to 1580\,nm, 
then gradually drops from 1580\,nm to 1620\,nm. 
The absolute DE is $=90.9\pm2.4\%$ at 1535\,nm.
Unfortunately, 
we were not able to measure DE at wavelengths 
shorter than 1510\,nm nor longer than 1620\,nm, 
because of a limited wavelength range of the probe light. \\

\begin{figure*}[hbt]
\centerline{\includegraphics[width=16cm]{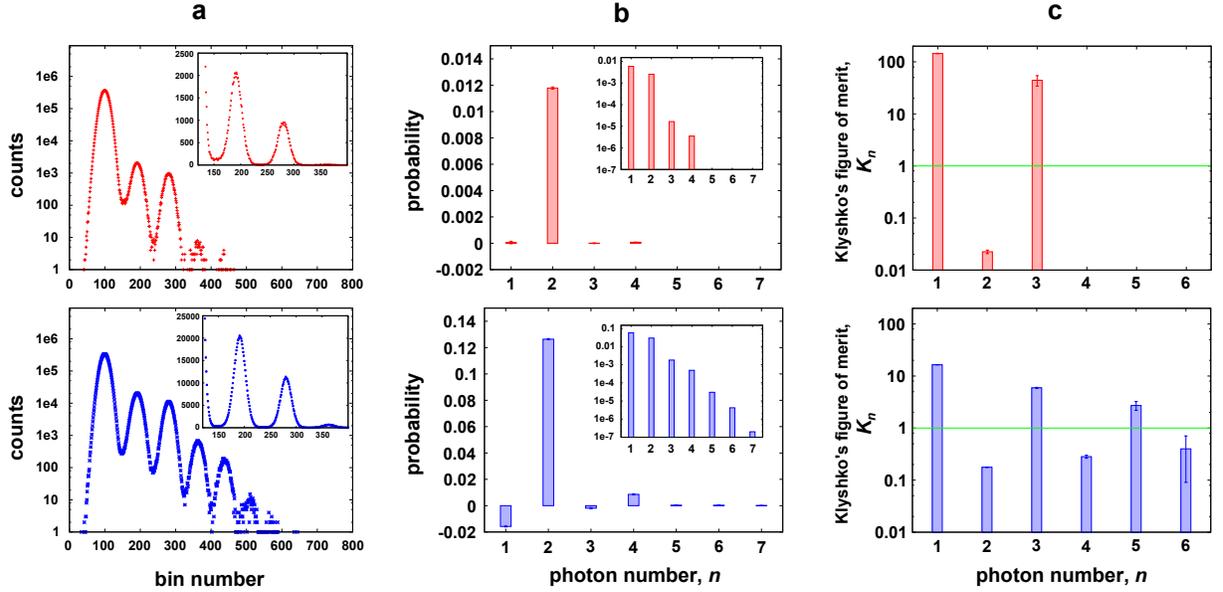}}
\caption{\textbf{Photon statistics of multimode squeezed states.} 
(\textbf{a}) 
Pulse height distributions from TES output waveforms 
for weak and strong pump conditions. 
The main graphs are in log scale, 
while the insets are in linear scale.
(\textbf{b}) 
Reconstructed photon probability distributions 
with the loss compensated. 
The insets are raw photon-probability distributions 
in log scale for the vertical axes.
(\textbf{c}) 
Klyshko's figure of merit in log scale.
All the error bars in (\textbf{b}) and (\textbf{c}) 
were statistical errors calculated 
by linear error-propagation method. 
We assume that photon-counting events obey Poissonian statistics.
}
\label{tes}
\end{figure*}

Figure \ref{tes} shows 
the experimental results for the multimode squeezed states. 
The red (blue) histograms correspond to 
a weak (strong) pumping condition in common.
We used a pump pulse energy of $\sim$1\,pJ ($\sim$10\,pJ) in the weak (strong) condition.
Figure \ref{tes}a shows 
the pulse height distribution of TES output waveforms. 
The total number of events was $1\times10^{7}$ for each data set. 
The vertical axes in the main graphs are numbers of counts shown in log scale, 
while those in the insets are in linear scale.
Every peak in each distribution was clearly separated, 
and was fitted with a Gaussian function 
in order to determine thresholds 
for calculating photon-number probabilities. 
The FWHM of the each Gaussian peak 
corresponds to an energy resolution and was calculated to be about 0.2\,eV,
compared with a photon energy ($\sim$0.8\,eV) of the monochromatic fundamental light at 1535\,nm.

Figure \ref{tes}b insets show 
the photon-probability distributions 
obtained from the pulse height distributions 
without any correction of losses. 
In each condition, 
the results clearly show super-Poissonian distribution,
whose feature could be confirmed by the Fano factors, 
which are greater than the unity,
yielding $1.462\pm0.002$ (red) and $1.494\pm0.001$ (blue), 
as well as by $g^{(2)}$ values,  
$42.746\pm0.195$ (red) and $4.978\pm0.008$ (blue).
In the weak pumping condition, 
two-photon generation is dominant in the squeezed vacuum state.
Then, the overall DE including the fiber-coupling efficiency 
can be evaluated by comparing the single and the two-photon probabilities, 
according to the formula, 
$\displaystyle\eta_{\mathrm{DE}} 
= \frac{2{P_2}/{P_1}}{1+2{P_2}/{P_1}}$ \cite{Waks04}.
The overall DE was thus estimated to be $46.2\pm 0.2\%$.

The main graphs in Fig. \ref{tes}b 
are reconstructed photon number statistics 
after compensating the loss using the relation 
$\displaystyle P_m 
= 
\sum_{k=m}^{M} \frac{k!}{m!(k-m)!} 
\eta_{\mathrm{DE}}{}^{m} (1-\eta_{\mathrm{DE}})^{k-m} p_k$,
where 
$p_k$ and $P_m$ denote ideal (lossless case) 
and experimentally measured photon-number probabilities, 
respectively \cite{Waks04}.
Dark counts were negligible compared to the average photon-number of measured photon statistics.
$M$ is a cutoff photon-number chosen to be large enough 
in order to prevent conversion errors by truncating higher photon-numbers.
We chose $M=20$ here.
The reconstructed statistics in red (upper panel) 
in the weak pumping condition 
only contains the two-photon component. 
On the other hand, 
the statistics in blue (lower panel) 
in the strong pumping condition 
contains the four-photon component apparently. 
Thus they exhibit even-rich photon-number statistics 
due to the pairwise photon generation of squeezing. 
Negative values appeared in single- and three-photon components 
in the lower panel (in blue histogram)  
may be due to several reasons including 
(1) conversion errors as stated in the ref. \cite{Waks04, Achilles03}, 
(2) slight degradation of $\eta_{\mathrm{DE}}$ 
in the strong pumping condition 
from that determined in the weak pumping one,
because the spatial-mode shift of the squeezed state 
would occur due to the strong pump
and thus would reduce the coupling efficiency into the optical fiber to the TES.
Further analysis is left open.

We also evaluated the nonclassicality 
using Klyshko's criterion: 
$\displaystyle K_n=(n+1)\frac{P_{n-1}P_{n+1}}{nP_n^2} < 1$ 
($n = 1, 2,...) $ \cite{Klyshko_original_96, Waks06}. 
In Fig. \ref{tes}c, 
the classical limit (green horizontal line) 
is clearly violated for even-photon numbers 
($n = $2, 4, and 6), 
while all the odd-photon numbers do not violate the limit. 
Ideally, 
squeezed source consists of even-photon numbers 
and therefore all the $K_n$ for odd numbers must go infinite.
There were in practice some contributions of odd-photon numbers 
caused by the finite DE, 
resulting in the finite $K_n$ values for odd photon-numbers.\\

\begin{figure*}[hbt]
\centerline{\includegraphics[width=15cm]{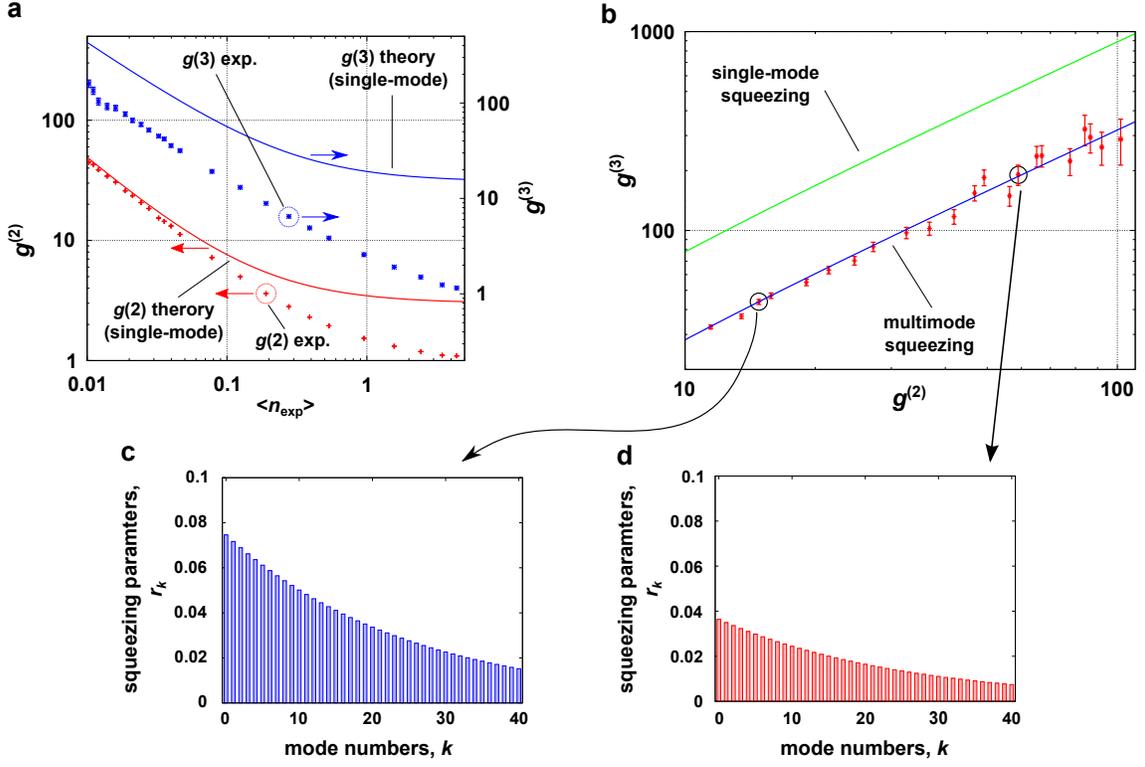}}
\caption{\textbf{Mode analysis of multimode squeezed states.}
(\textbf{a}) 
Experimental values of $g^{(2)}$ and $g^{(3)}$.
They show clear deviations from the single-mode cases (solid lines).
(\textbf{b}) 
$g^{(3)}$ values are plotted as a function of $g^{(2)}$ 
(red crosses).
The blue solid line is a result of fitting the data.
From a slope of the fitted curve, 
we obtained $\mu_\mathrm{exp}$, 
that determines a shape of the squeezer distribution.
The green solid line is the single-mode case, $\displaystyle g^{(3)} = 9 g^{(2)} - 12$, for reference.
(\textbf{c}) and (\textbf{d}) 
are reconstructed squeezing-parameter distributions 
using $\mu_{\mathrm{exp}}$ and $B_{\mathrm{exp}}$ 
for a pump pulse energy of $\sim$4\,pJ (blue) and $\sim$1\,pJ (red). 
}
\label{multimode}
\end{figure*}

Using the above PNRD data, 
we evaluate $g^{(2)}$ and $g^{(3)}$ values 
and analyze multimode structures of our squeezer
by applying the method in 
\cite{Christ11}. 
Figure \ref{multimode}a shows 
experimental $g^{(2)}$ and $g^{(3)}$ values
calculated from the photon number statistics 
using \textit{m}th factorial moment $n^{(m)}$ 
given by 
\begin{align}
g^{(m)} 
&= \frac{n^{(m)}}{\left<n_{\textrm{exp}}\right>^m}, 
\quad
(m=2 \,\textrm{or}\, 3)
\label{gm_exp}
\end{align}
where 
$\displaystyle n^{(m)} 
= \sum_{n=m}^{\infty}\frac{n!}{(n-m)!}P_n$ 
and 
$\displaystyle \left<n_{\textrm{exp}}\right>
=\sum_{n=0}^{\infty}nP_n$  
\cite{BR97}. 
The red and blue solid lines are 
theoretical plots for the single-mode squeezing 
given by 
\begin{align}
g^{(2)} = 3+\frac{1}{\left<n_{\textrm{exp}}\right>}, 
\quad
g^{(3)} = 15+\frac{9}{\left<n_{\textrm{exp}}\right>}. 
\end{align}
Because $g^{(m)}$ is a loss-tolerant measure, 
we use 
$\left<n_{\textrm{exp}}\right>=\eta_{\mathrm{DE}}\left<n\right>$ 
in order to compare the experimental $g^{(2)}$ and $g^{(3)}$ values 
to the theory.
There are clear deviations 
between the experimental and the theoretical values 
both for $g^{(2)}$ and $g^{(3)}$.
Note, $g^{(2)}$ values fitted well to the single-mode theory was reported in the ref. \cite{Gerrits11_SQZ}.

In the multimode case 
assuming a two-dimensional (2D) Gaussian JSD 
$f( \omega_{\mathrm{s}}, \omega_{\mathrm{i}} )$, 
mode weight distribution is thermal 
\cite{U'Ren_JSD_thermal_dist_03}. 
Then 
$g^{(2)}$ and $g^{(3)}$ can be expressed as follows 
\cite{Christ11} 
(also see Appendix \ref{AP2_multimode_g2g3}), 
\begin{align}
g^{(2)}_{\mathrm{multi}} & 
= 1 + 2 \mathcal{L}_4(\mu) + \frac{1}{B^2}, 
\label{g2_multi}\\
g^{(3)}_{\mathrm{multi}} & 
= 1 + 6 \mathcal{L}_4(\mu) + 8 \mathcal{L}_6(\mu) 
+ \frac{3 + 6 \mathcal{L}_4(\mu)}{B^2}, 
\label{g3_multi}
\end{align}
with 
$\displaystyle \mathcal{L}_4(\mu) = \frac{1-\mu^2}{1+\mu^2}$ 
and 
$\displaystyle \mathcal{L}_6(\mu) = \frac{(-1+\mu^2)^2}{1+\mu^2+\mu^4}$,  
where 
$B$ is an optical gain coefficient 
defined by 
$r_k=B\lambda_k$ with the normalized mode weight $\lambda_k$ 
($\sum_k \lambda_k^2=1$). 
The single parameter $\mu$ 
determines the thermal mode distribution as 
$\lambda_k = \sqrt{1-\mu^2}\mu^k$. 
The slope of 
$g^{(3)}_{\mathrm{multi}}$-vs-$ g^{(2)}_{\mathrm{multi}}$ 
can thus be defined as
\begin{eqnarray}
\mathcal{S}(\mu) \equiv 3 + 6 \mathcal{L}_4(\mu).
\label{slope}
\end{eqnarray}

The red crosses in Fig. \ref{multimode}b show 
the $g^{(3)}$ values as a function of $g^{(2)}$. 
By fitting the slope of $g^{(3)}$-vs-$g^{(2)}$ 
to $\mathcal{S}(\mu)$ 
using the linear least squares fitting technique, 
we derived 
$\mathcal{S}(\mu_{\textrm{exp}}) = 3.241\pm0.173$.
This corresponds to $\mu_{\textrm{exp}} = 0.961 \pm 0.028$. 
Once $\mu_{\textrm{exp}}$ is determined, 
each $B_\mathrm{exp}$ value can be obtained 
by eqs. (\ref{g2_multi}) and (\ref{g3_multi}) 
using corresponding
$g^{(2)}$ and $g^{(3)}$ values. 
Thus we can finally obtain the mode distribution 
of the squeezing parameter $r_k$ $(= B \lambda_k)$.
Reconstructed distributions up to the first 40 modes are shown 
in Fig. \ref{multimode}c and d for 
a pump pulse energy of $\sim$4\,pJ and $\sim$1\,pJ, respectively.
Corresponding optical gains $B_{\mathrm{exp}}$ 
were calculated to be $0.270$ (c) and $0.131$ (d), 
respectively.

The effective mode-number $K$ of the squeezed vacua contained in the reconstructed distribution
can be calculated using the relation $K=1/\displaystyle \mathcal{L}_4(\mu)$ \cite{Christ11},
yielding $K(\mu_{\textrm{exp}}) = 25$.
Furthermore, when the standard deviation of $\mu_{\textrm{exp}}$ is took into account,
$K$ is estimated to be within $14\sim90$.
This statistical uncertainty is a trade-off for probing the huge amount of modes with the finite PNRD events.
Hence, collecting more PNRD data should reduce the statistical uncertainty.
\\

In this work 
the nonclassical photon-number distributions 
as the even-odd photon-number oscillations 
were observed 
in an extreme broadband of 150\,nm FWHM. 
This bandwidth is, to our knowledge, 
the broadest one among those ever realized 
by any kinds of nonclassical light sources and 
by any types of photon detectors. 
The bands were technologically important ones, 
namely the telecom S-, C-, and L-band. 
The mode analysis showed that 
the several tens of orthonormal broadband modes 
were measured by our ultrabroadband PNR detector.
We anticipate our results and technique open up a new possibility 
to generate and characterize nonclassical light sources 
for a large-scale optical quantum network 
in the frequency domain \cite{Pysher11, Pinel12, Menicucci08}.

There are interesting open issues along with this work. 
Firstly 
more sophisticated methods for 
mode identification and characterization 
are needed to be developed. 
The method used here is just for evaluating 
the mode weight distributions 
under certain assumptions 
on the SPDC spectral correlation. 
More direct methods on mode analysis are desired.

Secondly
low-loss mode-separating devices 
should also be developed to implement 
prospective mode analyzing schemes 
and should be combined with highly efficient PNR detectors. 
In this regard, our setup can be extended into a phase-sensitive scheme 
by introducing phase coherent LOs. 
When a LO is seeded into the squeezer, 
squeezed coherent light can be generated, 
whose photon-number distributions exhibit 
unique phase dependent oscillation \cite{SW_ocsillation_87, Koashi_SubPoisson_93}. 
When, on the other hand, 
a LO is used at the detection side, 
then phase-sensitive PNR detection can be realized
\cite{Takeoka_Bin_Proj_Meas_06, Tsujino11, Izumi_PNRD_13, Becerra_q-receiver_4PSK_13}.
By appropriately engineering the LO modes, 
one can observe quantum states in designated modes 
in both schemes mentioned above. 
This would provide a first step toward 
realizing a quantum-mode analyzer. 

Finally broadband light sources play essential roles in diverse fields in the classical domain, 
such as high-capacity optical communications 
\cite{Ohara06}, 
optical coherence tomography 
\cite{Choi_Kurokawa2004_Frequency-comb-interferometer}, 
optical spectroscopy 
\cite{Coddington_Newbury2008_Coherent-multiheterodyne-spectroscopy}, 
and spectrograph calibration 
\cite{Ycas_Sigurdsson2012_spectrograph calibration}. 
Supercontinuum sources in the vicinity of the 1550\,nm 
telecommunication wavelength 
\cite{Genty_Dudley2007_Fiber-supercontinuum} 
are of particular importance 
for wavelength division multiplexing in fiber-optic communications 
\cite{Agrawal10}.  
Our ultrabroadband source and detection scheme shown here 
could also provide a way to utilize the multimode nonclassical states 
for realizing such capabilities in the quantum domain.

\acknowledgements
We thank R. Jin, G. Fujii, M. Fujiwara, T. Yamamoto, R. Shimizu, 
M. Koashi, T. Hirano and T. Gerrits for discussions,
and E. Sasaki and A. Sekine for technical support.
This work was supported by the 
Founding Program for World-Leading Innovative R\&D on Science and Technology
(FIRST) program.\\

\appendix

\section{Calibration of TES's detection efficiency depending on the wavelength.}
\label{AP1_DE_meas}
We used a continuous-wave (CW), wavelength-tunable ECDL (Santec, TSL-510) 
in order to measure the DE's wavelength dependence.
The CW light beam from the ECDL was connected to the fiber-coupler. 
One of the outputs from the coupler was guided to our TES after passing through the variable attenuator.
This power to the TES was heavily attenuated ($\sim$90\,dB) such that the average power was set to be below 1\,fW,
and was monitored via the other output port of the coupler using a power meter, whose wavelength sensitivity was precisely calibrated.
The coupler's splitting ratio and the degree of the attenuation were also precisely measured using the same power meter.
Because photon arrival was random using the CW source, 
total counts from the TES was collected by the threshold detector (SR400, Stanford Research).
Photon counts from the TES were a few kHz, and the raw waveforms were not overlapped each other.
We scanned the wavelength at each 3\,nm step from 1512 to 1620\,nm, and obtained a relative DE as shown in Fig. \ref{setup1}d.\\

\section{Second and third-order correlation functions 
for the multimode squeezer.} 
\label{AP2_multimode_g2g3}
In this multimode ``single-beam squeezer" case, 
the second and third intensity correlation functions 
are described in the ref. \cite{Christ11} as follows,
\begin{align}
g^{(2)}_{\mathrm{multi}} = 1 & + 2 \frac{\sum_k \sinh^4(r_k)}{[\sum_k \sinh^2(r_k)]^2}+\frac{1}{\sum_k\sinh^2(r_k)}
\label{g2_1}\\
g^{(3)}_{\mathrm{multi}} = 1 & + 6 \frac{\sum_k \sinh^4(r_k)}{[\sum_k \sinh^2(r_k)]^2} + 8 \frac{\sum_k \sinh^6(r_k)}{[\sum_k \sinh^2(r_k)]^3} 
\label{g3_1}\nonumber\\
& + \frac{3}{\sum_k\sinh^2(r_k)} + 6 \frac{\sum_k \sinh^4(r_k)}{[\sum_k \sinh^2(r_k)]^3} 
\end{align}
When a JSD takes a 2D Gaussian shape,
$\lambda_k$ forms a thermal distribution defined by a single parameter $\mu$, as $\lambda_k = \sqrt{1-\mu^2}\mu^k$.
When the $r_k$ is small enough ($r_k \ll 1$), 
that is valid in our experimental conditions,
the following approximations are valid as
\begin{eqnarray}
\frac{1}{\left[\sum_{k=0}^{\infty} \sinh^2(r_k)\right]^2}  
& \approx & \frac{1}{\left[\sum_{k=0}^{\infty} r_k^2\right]^2} \nonumber\\
& = & \frac{1}{B^2},\\
\frac{\sum_{k=0}^{\infty} \sinh^4(r_k)}{\left[\sum_{k=0}^{\infty} \sinh^2(r_k)\right]^2}  
& \approx & \frac{\sum_{k=0}^{\infty} r_k^4}{\left[\sum_{k=0}^{\infty} r_k^2\right]^2} 
=  \sum_{k=0}^{\infty} \lambda_k^4 \nonumber\\
& = & \mathcal{L}_4(\mu),\\
\frac{\sum_{k=0}^{\infty} \sinh^6(r_k)}{\left[\sum_{k=0}^{\infty} \sinh^2(r_k)\right]^3}  
& \approx & \frac{\sum_{k=0}^{\infty} r_k^6}{\left[\sum_{k=0}^{\infty} r_k^2\right]^3} 
= \sum_{k=0}^{\infty} \lambda_k^6 \nonumber\\
& = & \mathcal{L}_6(\mu).
\end{eqnarray}

Then, eqs. (\ref{g2_1}) and (\ref{g3_1}) can be simplified into the eqs. (\ref{g2_multi}) and (\ref{g3_multi}).
$g^{(2)}_{\mathrm{multi}}$ and $g^{(3)}_{\mathrm{multi}}$ values can be exactly calculated from the experimental photon statistics.
Furthermore, from eqs. (\ref{g2_multi}) and (\ref{g3_multi}),
$g^{(3)}_{\mathrm{multi}}$ can be expressed as a function of $g^{(2)}_{\mathrm{multi}}$ as
\begin{align}
g^{(3)}_{\mathrm{multi}} = & \left( 3 + 6 \mathcal{L}_4(\mu) \right) g^{(2)}_{\mathrm{multi}} \\\nonumber
& - 2 - 6 \mathcal{L}_4(\mu) - 12 \mathcal{L}_4{}(\mu)^2 + 8 \mathcal{L}_6(\mu).
\end{align}
The slope of $g^{(3)}_{\mathrm{multi}}$-vs-$ g^{(2)}_{\mathrm{multi}}$ is finally defined as eq. (\ref{slope}).\\

\section{Specification of PPKTP crystal and squeezing spectrum.}
\label{AP3_spectrum}

The JSD can be a product of two quantities as 
\begin{equation}
f(\omega_{\mathrm{s}},\, \omega_{\mathrm{i}}) 
= \alpha(\omega_{\mathrm{s}}, \, \omega_{\mathrm{i}}) 
\phi(\omega_{\mathrm{s}}, \, \omega_{\mathrm{i}})
\label{JSD}
\end{equation}
where 
$\alpha(\omega_{\mathrm{s}}, \, \omega_{\mathrm{i}}) $ 
describes pump-frequency envelop, 
and 
$\phi(\omega_{\mathrm{s}}, \, \omega_{\mathrm{i}})$ 
describes phase-matching function 
for the SPDC in a nonlinear crystal \cite{BK08}. 
Explicitly 
$\alpha(\omega_{\mathrm{s}}, \, \omega_{\mathrm{i}}) 
= \mathrm{sinc}(L \Delta k/2)$ 
with the crystal length $L$ 
and phase mismatch 
$\Delta k 
= 
k_{\mathrm{p}} - k_{\mathrm{s}} - k_{\mathrm{i}} - 2\pi/\Lambda$.
$\Lambda$ denotes a grating period of the PPKTP crystals 
and 24.2\,$\mu$m in our case.
Assuming that the transversal phase-matching is perfect, 
the longitudinal one can be described as 
$\Delta k 
= k_{\mathrm{p},z} - k_{\mathrm{s},z} - k_{\mathrm{i},z} 
- 2\pi/\Lambda$, 
because we used collinear configuration 
where pump, signal, and idler are all in the same polarization. 
Note that 
the beam propagating direction is taken to be along $z$-axis 
and $k_{\mu}(\omega_{\mu}) = \omega_{\mu} n_z(\omega_{\mu}) / c$ is the wave vectors. 
The frequencies are in a relation of 
$\omega_{\mathrm{i}} 
= \omega_{\mathrm{p}} - \omega_{\mathrm{s}} 
= 2\omega_{\mathrm{0}} - \omega_{\mathrm{s}}$. 
The corresponding Sellmeier equation for $n_z$ can be found in the ref. 
\cite{Fradkin99}.
The blue line in Fig. \ref{setup1}c was theoretically derived using the above formulas and parameters.

\end{document}